\documentclass[twocolumn,prb,showpacs,superscriptaddress,floatfix]{revtex4}
\usepackage{graphicx}
\usepackage{amsmath}
\usepackage{latexsym}
\usepackage{amssymb}
\usepackage{longtable}

\usepackage[citecolor=black,pagecolor=black,hyperfootnotes=false,hyperfigures=false,urlcolor=blue,colorlinks=true,linkcolor=black]{hyperref}

\begin{document}
\title{X-ray absorption and x-ray magnetic dichroism study on Ca$_{3}$CoRhO$_{6}$ and Ca$_{3}$FeRhO$_{6}$}
\author{T. Burnus}
\affiliation{II. Physikalisches Institut, Universit\"{a}t zu K\"{o}ln, Z\"{u}lpicher Str. 77, 50937 K\"{o}ln, Germany}
\author{Z. Hu}
\affiliation{II. Physikalisches Institut, Universit\"{a}t zu K\"{o}ln, Z\"{u}lpicher Str. 77, 50937 K\"{o}ln, Germany}
\author{Hua Wu}
\affiliation{II. Physikalisches Institut, Universit\"{a}t zu K\"{o}ln, Z\"{u}lpicher Str. 77, 50937 K\"{o}ln, Germany}
\author{J. C. Cezar}
\affiliation{European Synchrotron Radiation Facility, BP 220, 38043, Grenoble, France}
\author{S. Niitaka}
\affiliation{RIKEN, Institute of Physical and Chemical Research, 2-1, Hirosawa, Wako, Saitama 351-0198, Japan}
\affiliation{CREST, Japan Science and Technology Agency (JST), Kawaguchi, Saitama 332-0012, Japan}
\author{H. Takagi}
\affiliation{RIKEN, Institute of Physical and Chemical Research, 2-1, Hirosawa, Wako, Saitama 351-0198, Japan}
\affiliation{CREST, Japan Science and Technology Agency (JST), Kawaguchi, Saitama 332-0012, Japan}
\affiliation{Department of Advanced Materials Science, University of Tokyo, 5-1-5, Kashiwanoha, Kashiwa, Chiba 277-8581, Japan}
\author{C. F. Chang}
\affiliation{II. Physikalisches Institut, Universit\"{a}t zu K\"{o}ln, Z\"{u}lpicher Str. 77, 50937 K\"{o}ln, Germany}
\author{N. B. Brookes}
\affiliation{European Synchrotron Radiation Facility, BP 220, 38043, Grenoble, France}
\author{H.-J. Lin}
\affiliation{National Synchrotron Radiation Research Center, 101 Hsin-Ann Road, Hsinchu 30077, Taiwan}
\author{L. Y. Jang}
\affiliation{National Synchrotron Radiation Research Center, 101 Hsin-Ann Road, Hsinchu 30077, Taiwan}
\author{A. Tanaka}
\affiliation{Department of Quantum Matter, ADSM, Hiroshima University, Higashi-Hiroshima 739-8530, Japan}
\author{K. S. Liang}
\affiliation{National Synchrotron Radiation Research Center, 101 Hsin-Ann Road, Hsinchu 30077, Taiwan}
\author{C. T. Chen}
\affiliation{National Synchrotron Radiation Research Center, 101 Hsin-Ann Road, Hsinchu 30077, Taiwan}
\author{L. H. Tjeng}
\affiliation{II. Physikalisches Institut, Universit\"{a}t zu K\"{o}ln, Z\"{u}lpicher Str. 77, 50937 K\"{o}ln, Germany}

\date{\today}
\begin{abstract}
Using x-ray absorption spectroscopy at the Rh-$L_{2,3}$,
Co-$L_{2,3}$, and Fe-$L_{2,3}$ edges, we find a valence state of
Co$^{2+}$/Rh$^{4+}$ in Ca$_3$CoRhO$_6$ and of Fe$^{3+}$/Rh$^{3+}$
in Ca$_3$FeRhO$_6$. X-ray magnetic circular dichroism spectroscopy
at the Co-$L_{2,3}$ edge of Ca$_3$CoRhO$_6$ reveals a giant
orbital moment of about $1.7\mu_B$, which can be attributed to the
occupation of the minority-spin $d_0d_2$ orbital state of the
high-spin Co$^{2+}$ ($3d^7$) ions in trigonal prismatic
coordination. This active role of the spin-orbit coupling explains
the strong magnetocrystalline anisotropy and Ising-like magnetism
of Ca$_3$CoRhO$_6$.
\end{abstract}



\maketitle

\section{Introduction}

The quasi one-dimensional transition-metal oxides
Ca$_{3}AB$O$_{6}$ ($A=$ Fe, Co, Ni, \dots; $B=$ Co, Rh, Ir, \dots)
have attracted a lot of interest in recent years because of their
unique electronic and magnetic
properties.\cite{Fjellvaag1996,Aasland97,Kageyama1997,Kageyama1997b,Niitaka99,Maignan2000,Niitaka01,Niitaka01b,Martinez2001,Sampathkumaran02,Raquet02,Hardy03,Yao2007}
The structure of Ca$_{3}AB$O$_{6}$ contains one-dimensional (1D)
chains consisting of alternating face-sharing $A$O$_{6}$ trigonal
prisms and $B$O$_{6}$ octahedra. Each chain is surrounded by six
parallel neighboring chains forming a triangular lattice in the
basal plane. Peculiar magnetic and electronic behaviors are
expected to be related to geometric frustration in such a
triangle lattice with antiferromagnetic (AFM) interchain
interaction and Ising-like ferromagnetic (FM) intrachain coupling.
Ca$_{3}$Co$_{2}$O$_{6}$, which realizes such a situation, shows
stair-step jumps in the magnetization at regular intervals of the
applied magnetic field of $M_{s}/3$, suggesting ferrimagnetic spin
alignment. It has a saturation magnetization of $M_s = 4.8
\mu_{B}$ per formula unit at around 4 T.\cite{Maignan04} Studies
on the temperature and magnetic-field dependence of the
characteristic spin-relaxation time suggest quantum tunneling of
the magnetization similar to single-molecular
magnets.\cite{Hardy04} An applied magnetic field induces a large
negative magnetoresistance, apparently not related to the
three-dimensional magnetic ordering.\cite{Raquet02}
Band-structure calculations using the local-spin-density
approximation plus Hubbard U (LSDA+U) predicted that the
Co$^{3+}$ ion at the trigonal site, being in the high-spin (HS)
state ($S=2$), has a giant orbital moment of $1.57\mu_{B}$ due to
the occupation of minority-spin $d_2$ orbital, while the
Co$^{3+}$ ion at the octahedral site is in the low-spin (LS)
state ($S=0$).\cite{Wu05} An x-ray absorption and magnetic
circular dichroism study at the Co-$L_{2,3}$ edge has confirmed
this prediction.\cite{Burnus2006} Both studies explain well the
Ising nature of the magnetism of Ca$_3$Co$_2$O$_6$.

Ca$_{3}$CoRhO$_{6}$ and Ca$_{3}$FeRhO$_{6}$ have the same crystal
structure as Ca$_{3}$Co$_{2}$O$_{6}$, but different magnetic and
electronic properties: Neutron diffraction and magnetization
measurements also indicated intrachain-FM and interchain-AFM
interactions in Ca$_3$CoRhO$_6$ like in
Ca$_{3}$Co$_{2}$O$_{6}$.\cite{Niitaka01} In contrast,
susceptibility data on Ca$_{3}$FeRhO$_{6}$ reveal a single
transition into a three-dimensional AFM.\cite{Niitaka99,Davis03}
Although Ca$_{3}$CoRhO$_{6}$ has a similar magnetic structure as
Ca$_{3}$Co$_{2}$O$_{6}$, it exhibits considerable differences in
the characteristic temperatures in the magnetic susceptibility.
The high-temperature limit of the magnetic susceptibility shows a
Curie-Weiss behavior with a positive Weiss temperature of 150~K
for Ca$_{3}$CoRhO$_{6}$,\cite{Niitaka99} while 30~K was found for
Ca$_{3}$Co$_{2}$O$_{6}$.\cite{Aasland97,Kageyama1997} The
measured magnetic susceptibility undergoes two transitions at
$T_{c_1} = 90$~K and $T_{c_2}=25$~K for Ca$_{3}$CoRhO$_{6}$, and
at $T_{c_1} = 24$~K and $T_{c_2}=12$~K for
Ca$_{3}$Co$_{2}$O$_{6}$,\cite{Kageyama1997,Niitaka99,Niitaka01,Niitaka01b,Hardy03,Davis03}
which were attributed to FM-intrachain and AFM-interchain
coupling, respectively. In contrast, Ca$_{3}$FeRhO$_{6}$ has an
AFM ordering below $T_N=12$ K.\cite{Niitaka99,Davis03,Niitaka03}
Unlike Ca$_{3}$Co$_{2}$O$_{6}$, there is only one plateau at 4~T
and no saturation even at 18~T in the magnetization of
Ca$_{3}$CoRhO$_{6}$ at 70~K.\cite{Niitaka01} A partially
disordered state in Ca$_3$CoRhO$_6$ has been inferred by the
previous work of Niitaka \emph{et al.}\cite{Niitaka01b}

In order to understand the contrasting magnetic properties of
Ca$_{3}$CoRhO$_{6}$ and Ca$_{3}$FeRhO$_{6}$, and, particularly, the
type and origin of the intrachain magnetic coupling of these
quasi 1D systems, the valence, spin, and orbital states have to be
clarified. However, these issues have been contradictorily
discussed in previous theoretical and experimental studies. The
general-gradient-approximated (GGA) density-functional band
calculations\cite{Whangbo03} suggest a Co$^{3+}$/Rh$^{3+}$ state
in Ca$_{3}$CoRhO$_{6}$, while LSDA+U calculations with inclusion
of the spin-orbit coupling favor a Co$^{2+}$/Rh$^{4+}$ state and,
again, a giant orbital moment due to the occupation of
minority-spin $d_0$ and $d_2$ orbitals.\cite{Wu2007} Neutron
diffraction experiments on Ca$_{3}$CoRhO$_{6}$
\cite{Niitaka01b,Loewenhaupt} suggest the Co$^{3+}$/Rh$^{3+}$
state. However, based on the magnetic
susceptibility\cite{Niitaka99} and x-ray photoemission
spectroscopy\cite{Takubo05} the Co$^{2+}$/Rh$^{4+}$ state was
proposed. For Ca$_{3}$FeRhO$_{6}$, the Fe$^{2+}$/Rh$^{4+}$ state
was suggested in a magnetic susceptibility study,\cite{Niitaka99}
whereas M\"{o}ssbauer spectroscopy indicates a Fe$^{3+}$
state,\cite{Niitaka03} and thus Rh$^{3+}$.

In order to settle the above issues, in this work we first clarify
the valence state of the Rh, Co, and Fe ions in Ca$_{3}$CoRhO$_{6}$
and Ca$_{3}$FeRhO$_{6}$ using x-ray absorption spectroscopy (XAS)
at the $L_{2,3}$ edges of Rh, Co, and Fe. We reveal a valence
state of Co$^{2+}$/Rh$^{4+}$ in Ca$_3$CoRhO$_6$ and of
Fe$^{3+}$/Rh$^{3+}$ in Ca$_3$FeRhO$_6$. Then, we investigate the
orbital occupation and magnetic properties using x-ray magnetic
circular dichroism (XMCD) experiments at the Co-$L_{2,3}$ edge of
Ca$_3$ChRhO$_6$. We find a minority-spin $d_0d_2$ occupation for
the HS Co$^{2+}$ ground state and, thus, a giant orbital moment of
about $1.7\mu_B$. As will be seen below, our results account well
for the previous experiments.

\section{Experimental}

Polycrystalline samples were synthesized by a solid-state reaction
and characterized by x-ray diffraction to be single
phase.\cite{Niitaka99} The Rh-$L_{2,3}$ XAS spectra were measured
at the NSRRC 15B beamline in Taiwan, which is equipped with a
double-Si(111) crystal monochromator for photon energies above 2
keV. The photon-energy resolution at the Rh-$L_{2,3}$ edge
($h\nu\approx 3000$--3150 eV) was set to 0.6 eV. The
Fe-$L_{2,3}$ XAS spectrum of Ca$_{3}$FeRhO$_{6}$ was measured at
the NSRRC Dragon beamline with a photon-energy resolution of 0.25
eV. The main peak at 709~eV of the Fe-$L_3$ edge of single
crystalline Fe$_2$O$_3$ was used for energy calibration. The
Co-$L_{2,3}$ XAS and XMCD spectra of Ca$_{3}$CoRhO$_{6}$ were
recorded  at the ID8 beamline of ESRF in Grenoble with a
photon-energy resolution of 0.2 eV. The sharp peak at 777.8~eV of
the Co-$L_3$ edge of single crystalline CoO was used for energy
calibration. The Co-$L_{2,3}$ XMCD spectra were recorded in a
magnetic field of 5.5 T; the photons were close to fully
circularly polarized. The sample pellets were cleaved \textit{in
situ} in order to obtain a clean surface. The pressure was below
$5\times10^{-10}$~mbar during the measurements. All data were
recorded in total-electron-yield mode.

\section{XAS and valence state}

\begin{figure}
\includegraphics[width=0.48\textwidth]{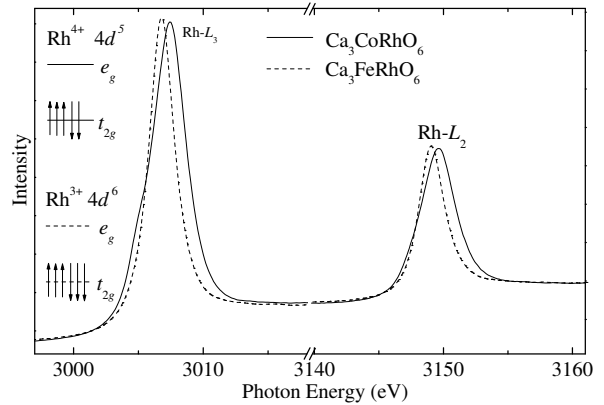}
\caption{\label{FIG:RhL23}The Rh-$L_{2,3}$ XAS spectra of
Ca$_{3}$CoRhO$_{6}$ and Ca$_{3}$FeRhO$_{6}$ and a schematic energy
level diagram for Rh$^{3+}$ $4d^6$ and Rh$^{4+}$ $4d^5$ configurations
in octahedral symmetry.
}\end{figure}

We first concentrate on the valence of the rhodium ions in both
studied compounds. For $4d$ transition-metal oxides, the XAS
spectrum at the $L_{2,3}$ edge reflects basically the unoccupied
$t_{2g}$- and $e_g$-related peaks in the $O_h$ symmetry. This is
due to the larger band-like character and the stronger
crystal-field interaction of the $4d$ states as well as due to the
weaker intra-atomic interactions as compared with $3d$
transition-metal oxides, where intra-atomic multiplet interactions
are dominant. The intra-atomic multiplet and spin-orbit
interactions in $4d$ elements only modify the relative intensity
of the $t_{2g}$- and $e_g$-related peaks. Fig. \ref{FIG:RhL23}
shows the XAS spectra at the Rh-$L_{2,3}$ edges of
Ca$_{3}$FeRhO$_{6}$ (dashed line) and Ca$_{3}$CoRhO$_{6}$ (solid
line). The Rh-$L_{2,3}$ spectrum shows a simple, single-peaked
structure at both Rh-$L_2$ and Rh-$L_3$ edges for
Ca$_{3}$FeRhO$_{6}$, while an additional low-energy shoulder is
observed for Ca$_{3}$CoRhO$_{6}$. Furthermore, the peak in the
Ca$_3$CoRhO$_6$ spectrum is shifted by 0.8~eV to higher energies
compared to that of the Ca$_3$FeRhO$_6$.

The single-peaked spectral structure for Ca$_{3}$FeRhO$_{6}$
indicates Rh$^{3+}$ ($4d^{6}$) with completely filled $t_{2g}$
orbitals, i.e. only transitions from the $2p$ core levels to the
$e_{g}$ states are possible. The results are in agreement with
M\"ossbauer spectroscopy.\cite{Niitaka03} The shift to higher energies
from Ca$_{3}$FeRhO$_{6}$ to Ca$_3$CoRhO$_6$ reflects the increase
in the Rh valence from Rh$^{3+}$ to Rh$^{4+}$ as we can learn
from previous studies on $4d$ transition-metal
compounds.\cite{deGroot94,Hu00,Hu02,Sahu2002} Furthermore, for
Ca$_3$CoRhO$_6$ the spectrum shows a weak low-energy shoulder,
which is weaker at the Rh-$L_2$ edge than at the Rh-$L_3$ edge.
This shoulder can be attributed to transitions from the $2p$ core
levels to the $t_{2g}$ state, reflecting a $4d^5$ configuration
with one hole at the $t_{2g}$ state. Such spectral features were
found earlier for Ru$^{3+}$ in
Ru(NH$_{4}$)$_{3}$Cl$_{6}$.\cite{deGroot94,Sham83} Detailed
calculations reveal that the multiplet and spin-orbit interactions
suppress the $t_{2g}$-related peak at the $L_2$ edge for a $4d^5$
configuration.\cite{deGroot94,Hu00,Hu02,Sahu2002} Thus, we find a
Rh$^{4+}$ ($4d^5$) state for Ca$_3$CoRhO$_6$. Having determined a
Rh$^{3+}$ state in Ca$_3$FeRhO$_6$ and a Rh$^{4+}$ state in
Ca$_3$CoRhO$_6$, we turn to the Fe-$L_{2,3}$ and the Co-$L_{2,3}$
XAS spectra to further confirm the Fe$^{3+}$ state and the
Co$^{2+}$ state, as expected for charge balance.

\begin{figure}
\includegraphics[width =0.46\textwidth]{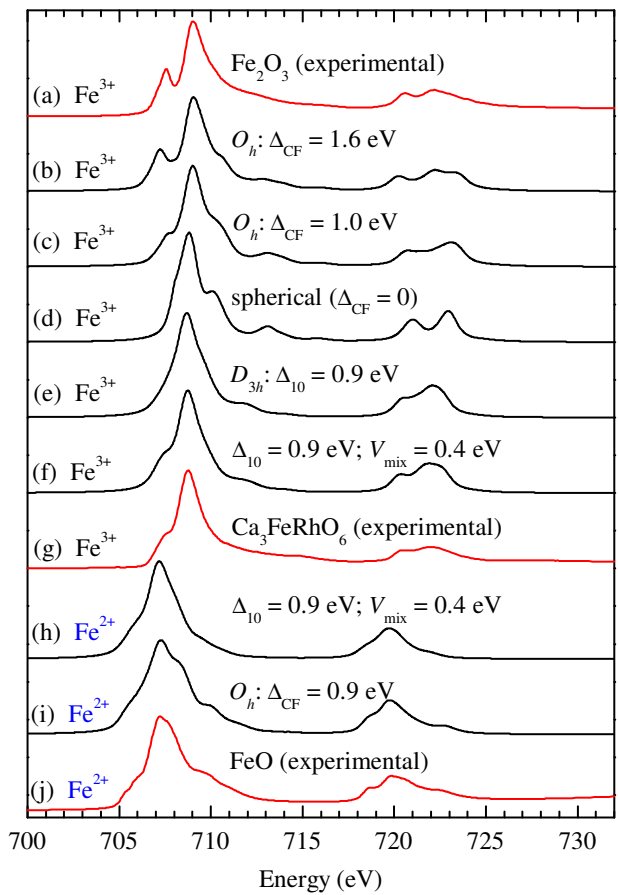}
\caption{\label{FIG:FeL23} (color online) Experimental XAS spectra
at the Fe-$L_{2,3}$ edge of (a) Fe$_2$O$_3$ (Fe$^{3+}$), (g)
Ca$_3$FeRhO$_6$, and  (j) FeO (Fe$^{2+}$), taken from
Park,\cite{Park94} together with simulated spectra (b, c) in
$O_h$, (d) spherical, and (e, f) $D_{3h}$ symmetry for Fe$^{3+}$
and simulated spectra in (h) $D_{3h}$ and (i)
$O_h$ symmetry for Fe$^{2+}$.}
\end{figure}

Figure \ref{FIG:FeL23} shows the experimental Fe-$L_{2,3}$ XAS
spectra of (g) Ca$_3$FeRhO$_6$, along with those of (a) single
crystalline Fe$_{2}$O$_{3}$ as a Fe$^{3+}$ reference and of (j)
FeO, taken from Ref. \onlinecite{Park94}, as a Fe$^{2+}$
reference. Additionally, calculated spectra for different
symmetries using purely ionic crystal-field multiplet
calculations \cite{deGroot90,Tanaka94,deGroot94,Thole97} are
shown. It is well known that an increase of the valence state of
the $3d$ transition-metal ion by one causes a shift of the XAS
$L_{2,3}$ spectra by about one eV towards higher
energies.\cite{Chen90,Mitra2003,Burnus2008} The main peak of the
Fe $L_3$ structure of the Ca$_{3}$FeRhO$_{6}$ lies 0.8 eV above
the main peak of the divalent reference FeO and only slightly
lower in energy than the one of Fe$_{2}$O$_{3}$ (Fe$^{3+}$). This
indicates trivalent iron ions in Ca$_3$FeRhO$_6$. The slightly
lower energy shift of Ca$_{3}$FeRhO$_{6}$ relative to
Fe$_{2}$O$_{3}$ can be attributed to the weak trigonal crystal field
in the former as compared an octahedral field in
the later, as we will show below.

The experimental spectra of the reference compounds, curve (a) for
Fe$_2$O$_3$ and curve (j) for FeO, can be well understood using
the multiplet calculations. For Fe$_2$O$_3$ we find a good
simulation taking a Fe$^{3+}$ ion in an octahedral symmetry with a
$t_{2g}$--$e_g$ splitting of 1.6 eV, which is depicted in curve
(b) in Fig. 2. For FeO, a good match with the experiment can be
found for the Fe$^{2+}$ in an octahedral environment with a
splitting of 0.9 eV, see curve (i). The weaker crystal field in
FeO, compared with Fe$_2$O$_3$, is consistent due to its larger
Fe--O bond length.

In order to understand the experimental Fe $L_{2,3}$ spectrum of
Ca$_3$FeRhO$_6$, we first return to the Fe$_2$O$_3$ spectrum.
When we reduce the $t_{2g}$--$e_g$ splitting from 1.6 eV (curve
b) via 1.0 eV (curve c) to 0.0 eV (curve d), we observe that the
the low-energy shoulder becomes washed out, while the high-energy
shoulder becomes more pronounced.\cite{deGroot90} Going further
to a trigonal crystal field, the high-energy
shoulder looses its intensity as shown in curve (e) for a
splitting of 0.9 eV between $d_{\pm1}$ ($d_{yz}$/$d_{zx}$)  and
$d_0$/$d_{\pm2}$ ($d_{3z^2-r^2}$/$d_{xy}$/$d_{x^2-y^2}$). The
experimental Fe-$L_{2,3}$ XAS spectrum of Ca$_3$FeRhO$_6$ in Fig.
\ref{FIG:FeL23}(g) can be well reproduced with this trigonal
crystal field of 0.9 eV and in addition a mixing parameter $V_{\rm
mix}= 0.4$ eV, which mixes the $d_{\pm2}$ with the $d_{\mp1}$
orbitals; the result for this Fe with the $3d^5$ high-spin
configuration is presented in curve (f).

We note that curve (f) has been generated with the Fe in the
trivalent state. As a check, we have also tried to fit the
experimental spectrum of Ca$_3$FeRhO$_6$ using a divalent Fe
ansatz. However, the simulation does not match, as is
illustrated in curve (h), in which we have used the same trigonal
crystal field splitting of 0.9 eV and mixing parameter of 0.4 eV.
To conclude, the Fe-$L_{2,3}$ and Rh-$L_{2,3}$ XAS spectra of
Ca$_{3}$FeRhO$_{6}$ firmly establish the Fe$^{3+}$/Rh$^{3+}$
scenario.

\begin{figure}
\includegraphics[width =0.44\textwidth]{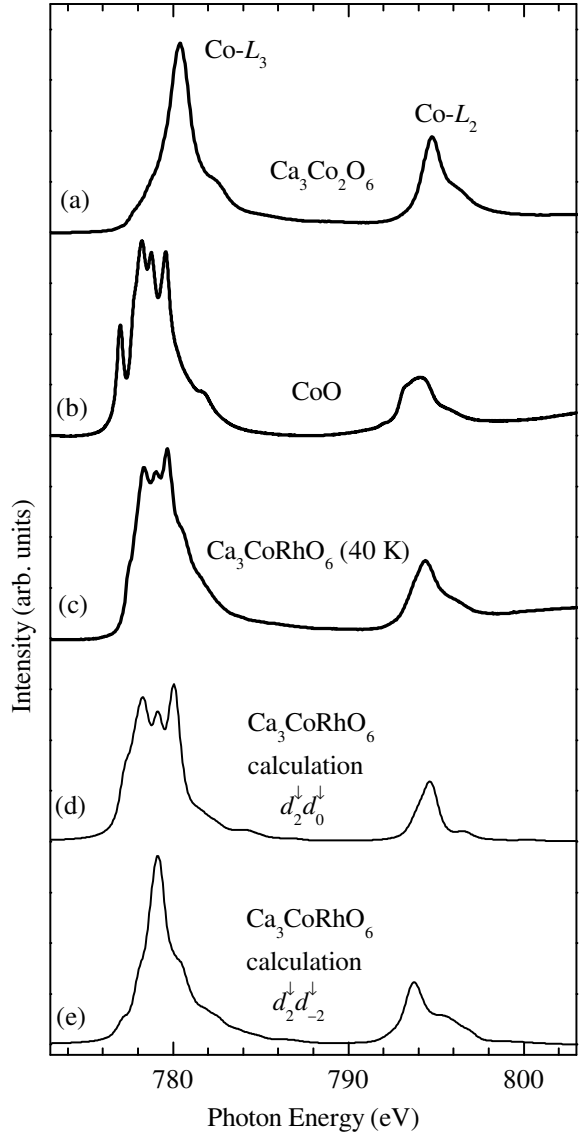}
\caption{\label{FIG:CoL23} The Co-$L_{2,3}$ spectra of (a)
Ca$_3$Co$_2$O$_6$ (Co$^{3+}$), (b) CoO (Co$^{2+}$), and (c)
Ca$_3$CoRhO$_6$. The simulated spectra of high-spin Co$^{2+}$ ($3d^7$) in
trigonal prismatic symmetry are shown in (d) for a
$d_0d_2$ and in (e) for a $d_2d_{-2}$ minority-spin orbital occupation.}
\end{figure}

For the Ca$_3$CoRhO$_6$ system, the Rh-$L_{2,3}$ XAS spectra
suggest that the Rh ions are tetravalent, implying that the Co
ions should be divalent. To confirm this Co$^{2+}$/Rh$^{4+}$
scenario we have to study explicitly the valence of the Co ion.
Fig. \ref{FIG:CoL23} shows the Co-$L_{2,3}$ XAS spectra of
Ca$_{3}$CoRhO$_{6}$ together with CoO as a Co$^{2+}$ and
Ca$_{3}$Co$_{2}$O$_{6}$ as a Co$^{3+}$ reference.\cite{Burnus2006}
Again we see a shift to higher energies from CoO to
Ca$_3$Co$_2$O$_6$ by about one eV. The Ca$_{3}$CoRhO$_{6}$
spectrum lies at the same energy position as the CoO spectrum
confirming the Co$^{2+}$/Rh$^{4+}$ scenario\cite{Wu2007} and
ruling out the Co$^{3+}$/Rh$^{3+}$ scenario.\cite{Whangbo03} The
result is fully consistent with the above finding from the
Rh-$L_{2,3}$ edge of Ca$_{3}$CoRhO$_{6}$ and in agreement with
previous results from x-ray photoemission
spectroscopy.\cite{Takubo05}

\begin{figure}
\includegraphics[width =0.46\textwidth]{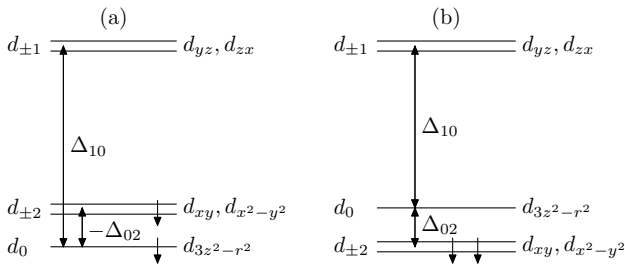}
\caption{\label{FIG:dLEVELS} Scheme of the two possible $3d$
occupations for a high-spin Co$^{2+}$ ion in trigonal prismatic symmetry,
ignoring the five up spins. (a) The $d_{0}d_2$ minority-spin occupation
allows for a large orbital magnetic moment, whereas (b) for
$d_2d_{-2}$ the orbital moment vanishes.}
\end{figure}

\section{XMCD and orbital occupation/moment}

After determining the valence states of Rh, Fe, and Co ions we
turn our attention to the orbital occupation and magnetic
properties of the Co$^{2+}$ ion at the trigonal-prism site. This
is motivated by the consideration that Co$^{2+}$ ions may have a
large orbital moment,\cite{Ghiringhelli} whose size depends on
details of the crystal field, while the high-spin Fe$^{3+}$
($3d^5$) and low-spin Rh$^{3+}$ ($4d^6$) ions in Ca$_3$FeRhO$_6$
have a closed subshell without orbital degrees of freedom and
thus no orbital moment.

In trigonal-prism symmetry the $3d$ orbitals are split into
$d_{\pm 1}$, $d_0$, and $d_{\pm 2}$ states, see Fig. \ref{FIG:dLEVELS}. In terms of
one-electron levels, the $d_{\pm 1}$ orbitals lie highest in
energy, while the lower lying $d_0$, and $d_{\pm 2}$ usually are
nearly degenerate. For a Co$^{3+}$ $d^6$ system, it is \textit{a
priori} not obvious from band structure calculations to say which
of these low lying orbitals gets occupied. Details, such as the
inclusion of the spin-orbit interaction, can become crucial.
Indeed, for Ca$_3$Co$_2$O$_6$, it was found from LDA+U
calculations \cite{Wu05} and confirmed by XMCD measurements
\cite{Burnus2006} that the spin-orbit interaction is crucial to
stabilize the occupation of the $d_2$ orbital, thereby
giving rise to giant orbital moments and Ising-type magnetism.
For a Co$^{2+}$ $d^7$ ion, however, the situation is quite
different. As we will explain below, the double occupation of the
$d_0d_2$ orbitals is energetically much more favored than that of
the $d_2d_{-2}$: the energy difference could be of order 1 eV
while the $d_0$ and $d_{\pm 2}$ by themselves could be degenerate
on a one-electron level. The consequences are straightforward:
the double occupation of $d_0d_2$, see Fig.~\ref{FIG:dLEVELS}(a),
should lead to a large orbital moment of $2\mu_{B}$ (neglecting
covalent effects) and Ising type of magnetism with the
magnetization direction fixed along the
chains.\cite{Niitaka01,Wu2007} In contrast, the $d_2d_{-2}$, see
Fig.~\ref{FIG:dLEVELS}(b), would have given a quenched orbital
moment.

\begin{figure}
\includegraphics[width =0.46\textwidth]{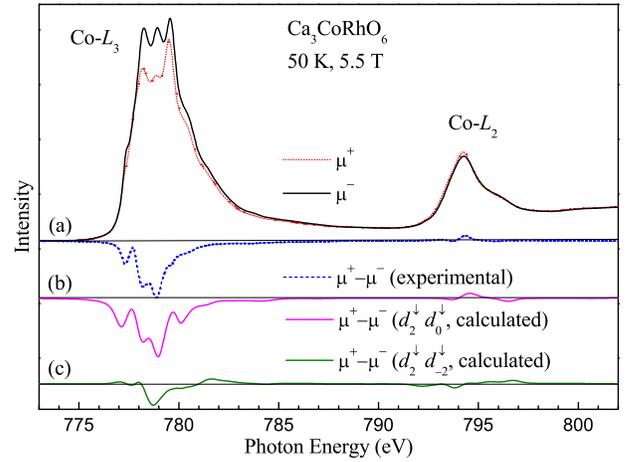}
\caption{\label{FIG:XMCDCo}(color online) (a) Measured soft x-ray
absorption spectra with parallel ($\mu^+$, red dotted curve) and
antiparallel ($\mu^-$, black solid curve) alignment between
photon spin and magnetic field, together with their difference
(XMCD) spectrum ($\mu^+-\mu^-$, blue dashed curve);
simulated XMCD spectra for (b) $d_0d_2$ (olive curve) and
(c) $d_2d_{-2}$ (magenta curve) minority-spin occupation of
the high-spin Co$^{2+}$.}
\end{figure}

In order to experimentally establish that the Co$^{2+}$ ion has
the $d_0d_2$ configuration, we have performed an XMCD study at
the Co-$L_{2,3}$ edges of Ca$_{3}$CoRhO$_{6}$. Fig.
\ref{FIG:XMCDCo} shows the Co-$L_{2,3}$ XMCD spectrum of
Ca$_{3}$CoRhO$_{6}$ taken at 50 K under 5.5 T. The spectra were
taken, respectively, with the photon spin parallel ($\mu^+$,
red dotted curve) and antiparallel ($\mu^-$, black solid curve)
to the magnetic field. One can clearly observe large differences
between the two spectra with the different alignments. Their
difference, $\mu^+-\mu^-$, is the XMCD spectrum (blue dashed
curve). An important feature of XMCD experiments is that there
are sum rules, developed by Thole and Carra \emph{et
al.},\cite{Thole92,Carra93} to determine the ratio between the
orbital ($m_{\rm orb} = L_z$) and spin ($m_{\rm spin} = 2S_z$)
contributions to the magnetic moment, namely

\begin{equation}
\label{EQ:MORBMSPIN} \frac{m_{\rm orb}}{m_{\rm spin}}
=\frac{2}{3}\frac{\Delta L_{3}+\Delta L_{2}}{\Delta L_{3}-2\Delta
L_{2}},
\end{equation}

\noindent here, $\Delta L_{3}$ and $\Delta L_{2}$ are the energy
integrals of the $L_{3}$ and $L_{2}$ XMCD intensity. The
advantage of these sum rules is that one needs not to do any
simulations of the spectra to obtain the desired quantum numbers.
In our particular case, we can immediately recognize the presence
of a large orbital moment in Fig. \ref{FIG:XMCDCo}(a), since there
is a large net (negative) integrated XMCD spectral weight. Using
Eq.~(\ref{EQ:MORBMSPIN}) we find $m_{\rm orb}/m_{\rm spin} =
0.63$. Since the Co$^{2+}$ ion is quite ionic, $m_{\rm spin}$ is
very close to the expected ionic value of $3\mu_{B}$. For
example, our LDA+U calculations yield $2.72 \mu_{B}$ for the
Co$^{2+}$ ion ($2.64 \mu_{B}$ for LDA) and Whangbo \emph{et al.}
obtained $2.71 \mu_{B}$ from GGA calculations.\cite{Whangbo03}
Using a value of $2.7\mu_B$ for the spin moment, we estimate
$m_{\rm orb}= 1.7\mu_{B}$, in nice agreement with our LDA+U
result of $1.69 \mu_{B}$, for the $d_0d_2$
minority-spin orbital occupation.\cite{Wu2007}

\begin{figure}
\includegraphics[width =0.46\textwidth]{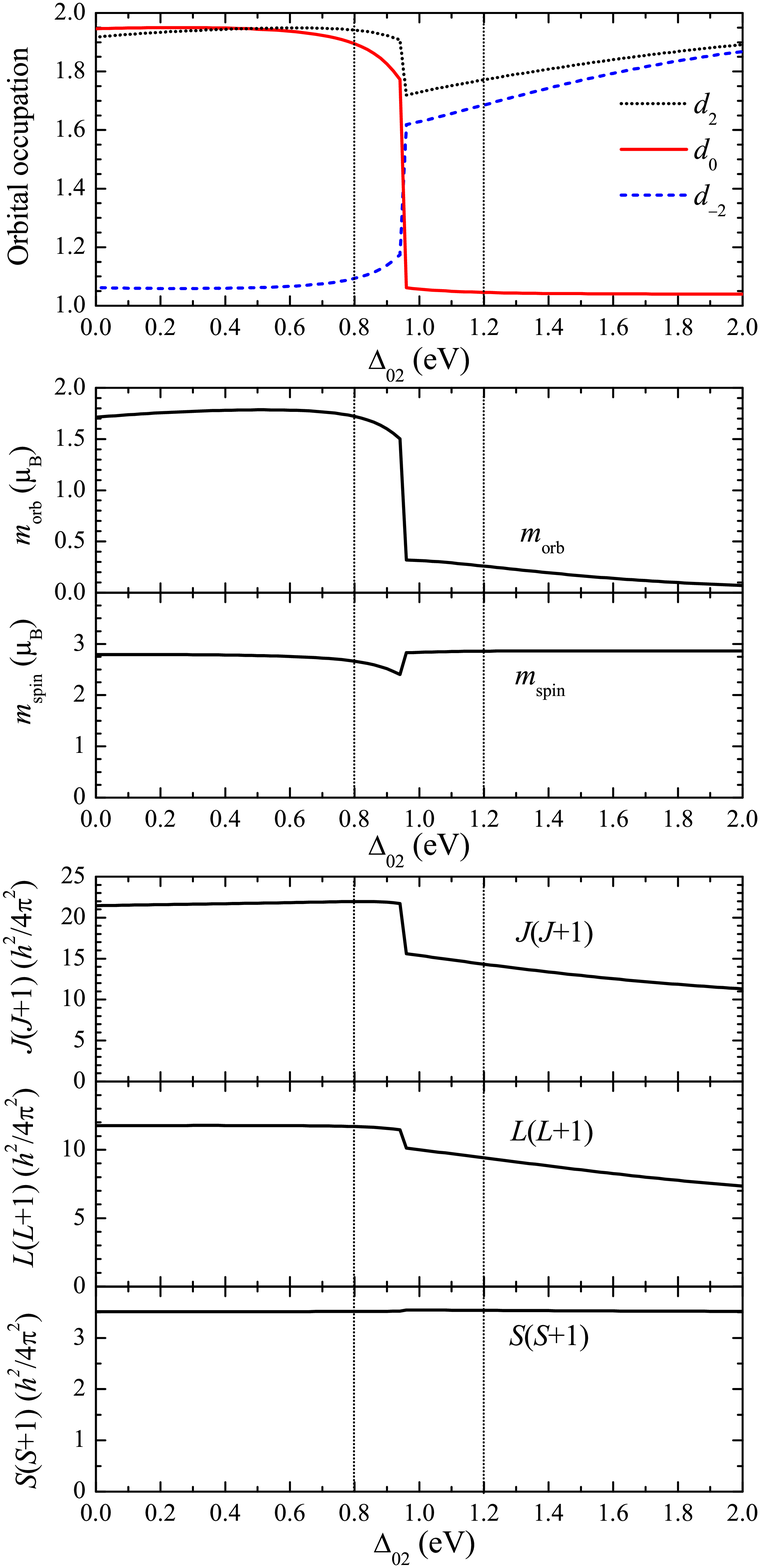}
\caption{\label{FIG:Moments}(color online) Top panel:
Occupation number of the $d_0$, $d_2$, and $d_{-2}$ orbitals as
function of the $d_0$--$d_{\pm2}$ splitting $\Delta_{02}$ [Fig. \ref{FIG:dLEVELS}(b)].
Middle panel: Orbital and spin moments ($m_{\rm orb}$ and $m_{\rm
spin}$) as function of $\Delta_{02}$. Bottom panel: $J(J+1)$,
$L(L+1)$, and $S(S+1)$ as function of $\Delta_{02}$.}
\end{figure}

To critically check our experimental and previous LDA+U
results\cite{Wu2007} regarding the $d_0d_2$ orbital occupation and
the giant orbital moment, we explicitly simulate the experimental
XMCD spectra using a charge-transfer configuration-interaction
cluster calculation,\cite{Tanaka94,deGroot94,Thole97} which
includes not only the full atomic multiplet theory and the local
effects of the solid, but also the oxygen $2p$--cobalt $3d$
hybridization. The results of the calculated Co-$L_{2,3}$ XAS and
XMCD spectra are presented in Figs. \ref{FIG:CoL23}(d) and
\ref{FIG:XMCDCo}(b), respectively. We can clearly observe that
the simulations reproduce the experimental spectra very well. The
parameters \cite{parameters} used are those which indeed give the
$d_0d_2$ orbital occupation for the ground state. The magnetic
quantum numbers found are $m_{\rm orb} = 1.65\mu_B$ and
$m_{\rm spin} = 2.46\mu_B$, yielding $m_{\rm orb}/m_{\rm
spin} =0.67$ and a total Co magnetic moment of $4.11\mu_B$.
With the Rh in the $S=1/2$ tetravalent state, the total
magnetic moment per formula unit should be around $5\mu_B$.
This is not inconsistent with the results of the high-field
magnetization study by Niitaka \emph{et al.}\cite{Niitaka01}:
they found a total moment of $4.05\mu_{B}$, but there the
saturation of the magnetization has not yet been reached even
under 18.7 Tesla. This can now be understood since the
magnetocrystalline anisotropy, associated with the active
spin-orbit coupling, is extremely strong and makes it difficult to
fully magnetize a powder sample as was used in their study.

We also have simulated the spectra for the $d_2d_{-2}$ scenario.
These are depicted in Figs. \ref{FIG:CoL23}(e) for the XAS and
\ref{FIG:XMCDCo}(c) for the XMCD. It is obvious that the
experimental spectra are not reproduced. The simulated line shapes
are very different from the experimental ones and the integral of
the simulated XMCD spectrum yields a vanishing orbital moment. We
therefore can safely conclude that the ground state of this
material is not $d_2d_{-2}$. For completeness we mention that the
magnetic quantum numbers found for this $d_2d_{-2}$
ansatz are $m_{\rm orb} = 0.03 \mu_{B}$ and $m_{\rm
spin} = 2.86 \mu_{B}$, yielding $m_{\rm orb}/m_{\rm spin}
=0.01$ and a total Co magnetic moment of $2.89 \mu_{B}$.

\section{Stability of the $d_2d_0$ state}

Having established that the ground state of Ca$_{3}$CoRhO$_{6}$
has the Co$^{2+}$ $d^7$ ion in the doubly occupied $d_0d_2$
orbital configuration and not in the $d_2d_{-2}$, it is
interesting to study its stability in more detail.  As already
mentioned above, for a Co$^{3+}$ $d^6$ ion, the $d_0$ and $d_{\pm
2}$ states can be energetically very close to each other. For a
Co$^{2+}$ $d^7$ ion, however, the $d_0d_2$ and $d_2d_{-2}$ states
are very much different in energy. This is illustrated in the top
panel of Fig. \ref{FIG:Moments}, in which we have calculated the
occupation numbers of the $d_0$, $d_2$, and $d_{-2}$ orbitals as a
function of $\Delta_{02}$, the one-electron level splitting
between the $d_0$ and $d_{\pm 2}$ orbitals. The $d_0d_2$ ground
state which gives the best simulation to the experimental XAS and
XMCD spectra was obtained with $\Delta_{02}\approx 0.4$ eV.
We can observe that the $d_0d_2$ situation is quite stable for a
wide range of $\Delta_{02}$ values, certainly up to 0.8 eV. With
a transition region between $\Delta_{02}=0.8$--1.2 eV, we find a
stable $d_2d_{-2}$ situation only for $\Delta_{02}$ values larger
than 1.2 eV. (For the $d_2d_{-2}$ simulations above we have used
$\Delta_{02} = 1.4$ eV.) This is a very large number indeed, and
it can be traced back to the multiplet character of the on-site
Coulomb interactions: an occupation of $d_2d_{-2}$ means a much
stronger overlap of the electron clouds as compared to the case
for a $d_0d_2$. This results in a higher repulsion energy, which
is not a small quantity in view of the atomic-like values of the
$F^2$ and $F^4$ Slater integrals determining the multiplet
splitting.\cite{Tanaka94,Antonides77}

In the middle panel of Fig. \ref{FIG:Moments} we also show the
expectation values for $m_{\rm orb}$ and $m_{\rm spin}$ when
varying $\Delta_{02}$. Again we clearly observe that the large
orbital-moment situation is quite stable.
To quench the orbital moment one would need much higher
$\Delta_{02}$ values. Important
is that the spin state does not change here. Bottom panel of Fig.
\ref{FIG:Moments} demonstrates that the high-spin state of the
Co$^{2+}$ ion is not affected by $\Delta_{02}$: the expectation
value $\langle S^2\rangle$ remains constant throughout at a value
consistent with an essentially $S=3/2$ state. Obviously, the $L^2$
and $J^2$ quantum numbers are strongly affected by $\Delta_{02}$.

\section{Conclusion}
To summarize, the Rh-$L_{2,3}$, Co-$L_{2,3}$ and Fe-$L_{2,3}$ XAS
measurements indicate Co$^{2+}$/Rh$^{4+}$ in Ca$_{3}$CoRhO$_{6}$
and Fe$^{3+}$/Rh$^{3+}$ in Ca$_{3}$FeRhO$_{6}$. The magnetic
properties of Ca$_3$FeRhO$_6$ are relatively simple as both the
HS Fe$^{3+}$ and LS Rh$^{3+}$ ions have a closed subshell and
thus no orbital degrees of freedom and no orbital moment. The
weak intrachain AFM coupling between the HS Fe ions can be
understood in terms of the normal superexchange via the
intermediate non-magnetic O--Rh--O complex. For Ca$_3$CoRhO$_6$,
the combined experimental and theoretical study of the
Co-$L_{2,3}$ XAS and XMCD spectra reveals a giant orbital moment
of about $1.7\mu_B$. This large orbital moment is connected with
the minority-spin $d_0d_2$ occupation for HS Co$^{2+}$ (3d$^7$)
ions in the peculiar trigonal prismatic coordination. The high FM
ordering temperature in Ca$_{3}$CoRhO$_{6}$, compared with that
of Ca$_{3}$Co$_2$O$_{6}$, can be attributed to the distinct
octahedral sites (which mediate the Co--Co magnetic coupling):
the magnetic Rh$^{4+}$ ion ($S = 1/2$) in the former and the
nonmagnetic Co$^{3+}$ ion ($S = 0$) in the latter.

We would like to thank Lucie Hamdan for her skillful technical and
organizational assistance in preparing the experiment. The
research in K\"oln is supported by the Deutsche
Forschungsgemeinschaft through SFB 608.


\begin{thebibliography}{40}
\bibitem[{par()}]{parameters}
\bibinfo{note}{Parameters (in eV). $U_{3d,3d}=5$, $U_{2\underline p,3d} = 6.5$,
  $\Delta = 4$, $\Delta_{10}^{\rm ionic} = 0.65$, $V_{\rm mix}^{\rm
  ionic}=-0.2$, $V_{pd\sigma}=-1.024$, $H_{\rm ex}=0.045$. The Slater integrals
  were reduced to 80\% of their Hartree-Fock value. $\Delta_{02}^{\rm
  ionic}=0.4$ ($d_0d_2$ scenario), $\Delta_{02}^{\rm ionic}=1.4$
  ($d_2d_{-2}$)}.
\end{thebibliography}
\end{document}